\documentclass[pre,aps,twocolumn,longbibliography]{revtex4-1}

\usepackage{hyperref}

\usepackage{amsmath,amsfonts,amssymb}
\usepackage{bm}
\usepackage{graphicx}
\usepackage[usenames]{color}

\def\cal#1{\mathcal{#1}}
\def\eqq#1{Eq.~(\ref{#1})}

\def\eq#1{(\ref{#1})}

\def\f#1{Fig.~\ref{#1}}
\def\ff#1{Figure~\ref{#1}}

\def\t#1{Table~\ref{#1}}

\def\s#1{Section~\ref{#1}}

\def\c#1{~\cite{#1}}

\def\th{\bm \theta}
\def\e{{\rm e}}
\def\d{{\rm d}}
\def\ce{\cal C}

\def\av#1{\langle #1 \rangle}

\def\0{\texttt{000}}
\def\1{\texttt{001}}
\def\2{\texttt{010}}
\def\3{\texttt{011}}
\def\4{\texttt{100}}
\def\5{\texttt{101}}
\def\6{\texttt{110}}
\def\7{\texttt{111}}

\def\beq{\begin{equation}}
\def\eeq{\end{equation}}
\def\bea{\begin{eqnarray}}
\def\eea{\end{eqnarray}}

\begin{document}

\title{Learning stochastic dynamics and predicting emergent behavior using transformers}
\author{Corneel Casert}
\email{{corneel.casert@ugent.be}}
\affiliation{Department of Physics and Astronomy, Ghent University, 9000 Ghent, Belgium}
\author{Isaac Tamblyn}
\email{{isaac.tamblyn@uottawa.ca}}
\affiliation{University of Ottawa, Ottawa, ON K1N 6N5, Canada}
\affiliation{Vector Institute for Artificial Intelligence, Toronto, ON M5G 1M1, Canada}
\author{Stephen Whitelam}
\email{{swhitelam@lbl.gov}} 
\affiliation{Molecular Foundry, Lawrence Berkeley National Laboratory, 1 Cyclotron Road, Berkeley, CA 94720, USA}

\begin{abstract}

We show that a neural network originally designed for language processing can learn the dynamical rules of a stochastic system by observation of a single dynamical trajectory of the system, and can accurately predict its emergent behavior under conditions not observed during training. We consider a lattice model of active matter undergoing continuous-time Monte Carlo dynamics, simulated at a density at which its steady state comprises small, dispersed clusters. We train a neural network called a transformer on a single trajectory of the model. The transformer, which we show has the capacity to represent dynamical rules that are numerous and nonlocal, learns that the dynamics of this model consists of a small number of processes. Forward-propagated trajectories of the trained transformer, at densities not encountered during training, exhibit motility-induced phase separation and so predict the existence of a nonequilibrium phase transition. Transformers have the flexibility to learn dynamical rules from observation without explicit enumeration of rates or coarse-graining of configuration space, and so the procedure used here can be applied to a wide range of physical systems, including those with large and complex dynamical generators.

\end{abstract}

\maketitle

{\em Introduction---}  Learning the dynamics governing a simulation or experiment is a demanding task because the number of possible dynamical transitions increases exponentially with the physical size of the system. For large systems these transitions are too numerous be enumerated explicitly. Here we show that it is possible to circumvent this restriction by using a neural network to parameterize a stochastic dynamics. In particular, we show that a recently-introduced neural network called a transformer\c{vaswani2017attention}, popular in the fields of natural-language processing and computer vision\c{devlin2018bert,radford2019language,brown2020language,parmar2018image,dosovitskiy2020vit}, can express a general dynamics in an efficient way. It can be trained offline, i.e. by observation only\c{levine2020offline}, to learn the dynamical rules of a model, even when those rules are numerous and nonlocal. Forward-propagated trajectories of the trained transformer can then be used to reproduce the behavior of the observed model, and to predict its behavior when applied to conditions not seen during training.

Previous work has shown that it is possible to learn the rules of deterministic dynamics, such as deterministic cellular automata\c{wulff1992learning,gilpin2019cellular,grattarola2021learning}, or of stochastic dynamics for small state spaces, using maximum-likelihood estimation on the rates of the generator\c{mcgibbon2015efficient}. Similar methods have been used to learn the form of intermolecular potentials that influence the dynamical trajectories of particles undergoing Brownian motion\c{frishman2020learning,garcia2018high,chen2021maximum}. Several approaches exist in which high-dimensional dynamical systems are approximated by lower-dimensional ones, such as Markov-state models\c{prinz2011markov, bowman2013introduction}. In some cases the coarse-graining procedures used to produce such models involve variational methods\c{wu2020variational} and neural networks\c{mardt2018vampnets}. Coarse-graining methods have also been used to obtain deterministic hydrodynamic equations from stochastic trajectories of active matter, allowing for the extraction of hydrodynamic transport coefficients\c{supekar2021learning,maddu2022learning}. Our work complements these approaches by showing that it is possible to learn the dynamical rules of stochastic systems without explicit enumeration of rates or coarse-graining of configuration space, thereby allowing treatment of large and complex systems. From the observation of a single dynamical trajectory a transformer can identify how many classes of process exist and what are their rates, providing physical insight into the dynamics and allowing it to be simulated in new settings, where new phenomena can be discovered.

We focus on the case of a lattice model of active matter, simulated using continuous-time Monte Carlo dynamics\c{lattice1} (in the Supplemental Information we show that the transformer can be used to treat a second class of model, one realization of which has nonlocal dynamical rules). We allow the transformer to know that the rates for this dynamics are independent of time, and that possible moves consist of single particles rotating in place or translating one lattice site at a time (both restrictions can be relaxed within our framework). However, we do not allow the transformer to know the rates for each move, and, because each rate could in principle depend on the state of the entire system, explicit enumeration of rates would require a generator with many more than $10^{100}$ entries for the system size considered. From observation of a single trajectory of the model, carried out at a density at which its steady state comprises small, dispersed clusters, the transformer learns that particle moves fall into a small number of classes, and accurately determines the associated rates. Forward-propagated trajectories of the trained transformer at the training density reproduce the model's behavior. Moreover, forward-propagated trajectories of the transformer carried out at densities higher than that used in training exhibit motility-induced phase separation\c{gonnella2015motility,Cates_2015,o2021introduction,Redner_2016,omar2021phase}. The details of this phase separation match those of the original model, although that information was not available to the transformer during training.

The trained transformer is therefore able to accurately extrapolate a learned dynamics to predict the existence and details of an emergent phenomenon that it had not previously observed. Given that the transformer is expressive enough to represent a nonlocal dynamics, these results indicate the potential of such devices to learn dynamical rules and study emergent phenomena from observations of dynamical trajectories in a wide variety of settings.

{\em Learning dynamics by observation ---} Imagine that we are given a dynamical trajectory $\omega$ of total time $T$. The trajectory starts in configuration (microstate) $\ce_0$, and visits $K$ additional configurations $\ce_k$ (\f{fig:schem}(a)). In configuration $\ce_k$ it is resident for time $\Delta t_{C_k}$. Schematically, 
\beq
\omega = \ce_0 \xrightarrow[]{\Delta t_{\ce_0}} \ce_1 \xrightarrow[]{\Delta t_{\ce_1}} \cdots\ce_{K-1} \xrightarrow[]{\Delta t_{\ce_{K-1}}}\ce_K \xrightarrow[]{\Delta t_{K}} \ce_K, \nonumber
\eeq
where $\Delta t_K \equiv T-\sum_{k=0}^{K-1} \Delta t_{\ce_k}$. We are told that $\omega$ was generated by a dynamics whose rates $W^{\star}_{\ce \to \ce' }$ for passing between configurations $\ce$ and $\ce'$ we do not know. We will call this unknown dynamics the {\em original} dynamics. 

Here we show it is possible to efficiently learn the original dynamics offline, i.e. solely by observation of $\omega$. We start by constructing a {\em synthetic} dynamics, which consists of a set of allowed configuration changes $\{\ce \to \ce'\}$, which must include those observed in $\omega$, and associated rates $W^{(\th)}_{\ce \to \ce' }$. Without prior knowledge of the system we should allow the rates for these moves to depend, in principle, on the entire configuration of the system. The number of possible rates grows exponentially with system size, and so treating a system of appreciable size requires the use of an expressive parameterization of the synthetic dynamics. Here we parameterize the rates $W^{(\th)}_{\ce \to \ce' }$ of the synthetic dynamics using the weights $\th$ of a neural network.

\begin{figure}[] 
\includegraphics[width=1.02\columnwidth]{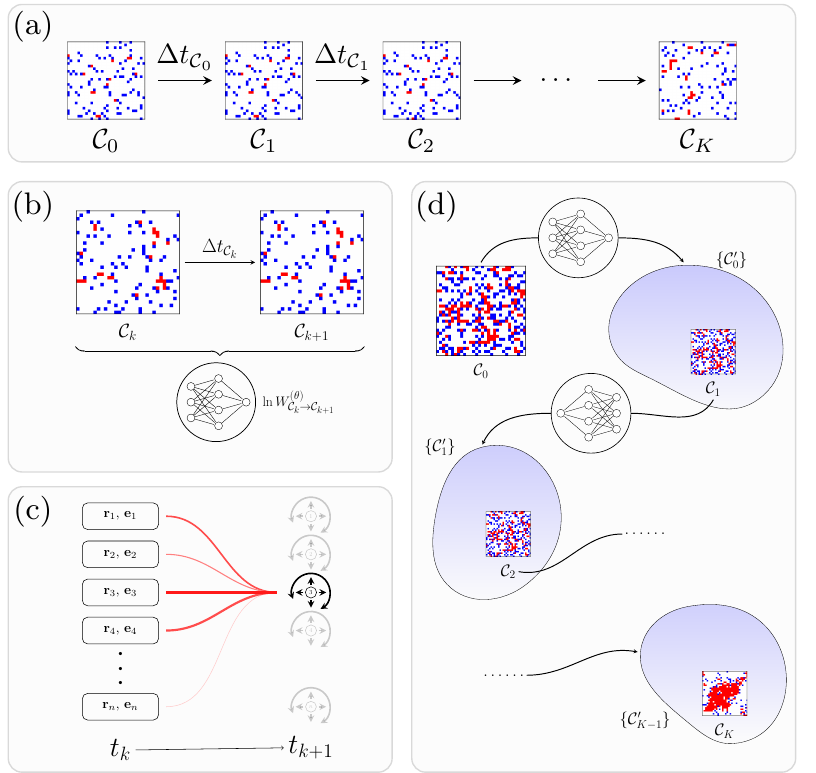} 
   \caption{\label{fig:schem} Schematic of our dynamics-learning procedure. (a) We are provided with a trajectory $\omega$, a time series of configurations, and wish to learn the dynamics that created it. For the lattice-based active-matter model studied here, red or blue indicates a particles whose orientation vector points toward an occupied or empty site, respectively. (b) We parameterize a general dynamics using a neural network called a transformer. Rates connecting configurations depend on the weights of the transformer, which are adjusted during training in order to maximize the log-likelihood with which it would have generated $\omega$. (c) The transformer receives the position and orientation of all particles, and must calculate the transition rates to translate or rotate each particle. To do so, it must learn which interactions affect these rates (line thickness denotes attention given to each particle), and their numerical values. (d) Once trained, the neural-network dynamics can be forward-propagated to generate new trajectories, even under conditions not observed in $\omega$. The transformer calculates the rates for all possible transitions $\mathcal{C}_k \to \{\mathcal{C}_k^\prime\}$, represented by the blue blobs, at each step.}
\end{figure}

\begin{figure}[] 
\includegraphics[width=\columnwidth]{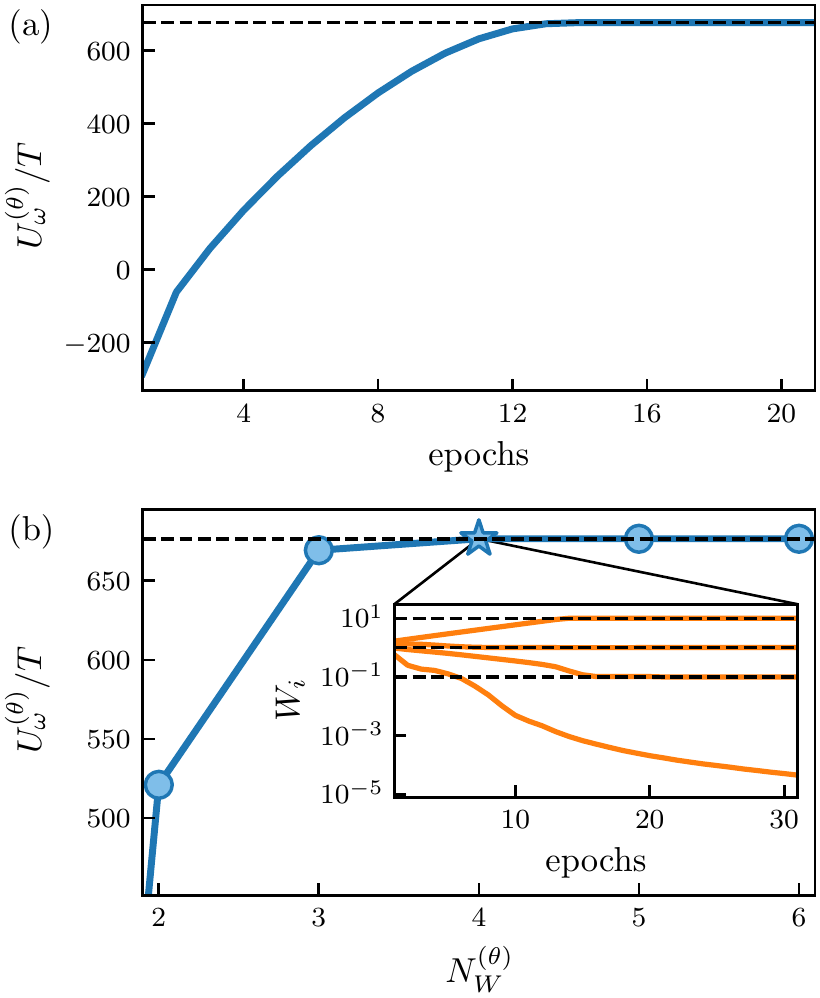} 
   \caption{\label{fig:ll} (a) Training of a transformer in Mode 1 (unrestricted rates) to maximize the log-likelihood $U_\omega^{(\th)}$, \eqq{phi}, of the training trajectory $\omega$. The horizontal black line denotes the value of the path weight associated with the original model. (b) Dependence of $U_\omega^{(\th)}$ for a transformer trained in Mode 2 on $N_W^{(\th)}$, the number of distinct classes of move it was asked to identify. This procedure allows us to identify the existence of $N_W^{\star} =4$ distinct rates. Inset: Evolution of the rates during training in Mode 2, with $N_W^{(\th)} = 4$. The horizontal black lines denote the values of the rates in the original dynamics.}
\end{figure}

 \begin{figure}[] 
\includegraphics[width=\columnwidth]{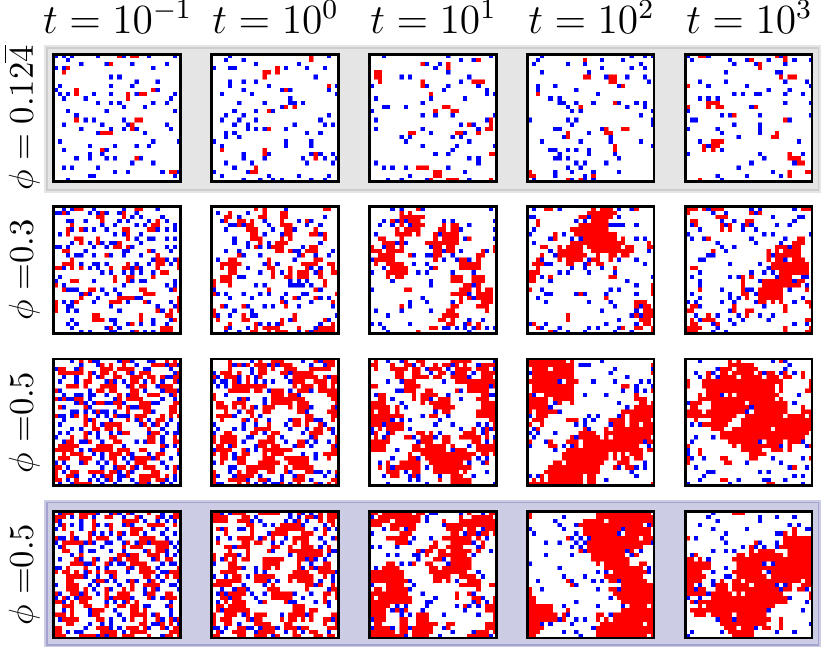} 
   \caption{\label{fig:trajs_phi} Trajectories of the lattice active-matter model generated using the dynamics learned by the transformer. The top row shows time-ordered snapshots of a trajectory generated at density $\phi = 0.12\overline{4}$, the value used during training. The two middle rows use densities $\phi = 0.3$ and $\phi = 0.5$; here, motility-induced phase separation can be seen. For comparison, the bottom row shows a trajectory generated with the original dynamics at $\phi = 0.5$. }
\end{figure}

\begin{figure*}[] 
\includegraphics[width=\textwidth]{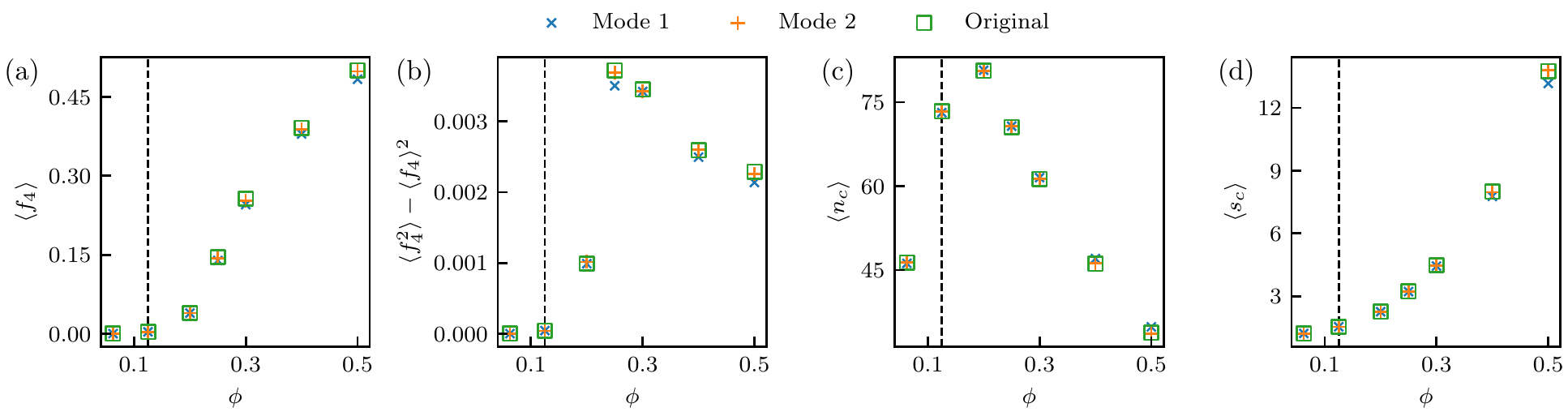} 
  \caption{\label{fig:obs} Comparison of learned and original dynamics. (a) Time-averaged fraction of particles with four neighboring occupied sites, $f_4$, as a function of density $\phi$, measured over trajectories generated using a transformer trained in Mode 1 (crosses) and Mode 2 (plusses). Angle brackets denote time averages. For each density, 10 trajectories of length $T=10^4$ were created using each of the dynamics. Error bars are smaller than the symbol size.  Training was done only at $\phi =  0.12\overline{4}$ (vertical dashed line). Squares denote results obtained using the original dynamics. The remaining panels have the same format and show (b) the variance of $f_4$, (c) the number of clusters $n_c$, and (d) the averaged cluster size $s_c$.}
\end{figure*}

One way to learn the original dynamics is to propagate the synthetic dynamics and alter its parameters $\th$ until the dynamical trajectories it generates resemble $\omega$. One drawback of this approach is that original and synthetic dynamics are stochastic, and so comparison of trajectories can be made only in a statistical sense, potentially requiring the generation of many synthetic trajectories at each stage of training. In addition, a comparison of this nature would require the introduction of additional order parameters, different combinations of which may result in different outcomes of training. Instead, we train the synthetic dynamics by maximizing the log-likelihood $U_\omega^{(\th)}$ with which it would have generated $\omega$\c{mcgibbon2015efficient}. We consider continuous-time Monte Carlo dynamics, in which case
 \bea
\label{phi}
U_\omega^{(\th)}&=&\sum_{k=0}^{K-1} \left(\ln W_{\ce_k \to \ce_{k+1}}^{(\th)} -\Delta t_{\ce_k} R_{\ce_k}^{(\th)}\right) \nonumber \\
&-& \Delta t_K R_{\ce_K}^{(\th)};
\eea
see \s{ctmc}. Training proceeds by adjusting the parameters $\th$ of the neural network until $U_\omega^{(\th)}$ no longer increases; see \f{fig:schem}(b). The synthetic dynamics obtained in this way -- the {\em learned} dynamics -- is then the best approximation to the original dynamics that our choice of allowed configuration changes and method of training allows: for a sufficiently long trajectory $\omega$, the dynamics that maximizes $U_\omega^{(\th)}$ is the original dynamics, $W^{\star}_{\ce \to \ce' }$. Other types of dynamics, such as Langevin dynamics, can be studied analogously, by choosing an appropriate replacement for the trajectory log-likelihood.

{\em Original dynamics---} The original dynamics we consider is a lattice model of active matter simulated using continuous-time Monte Carlo\c{lattice1}. It consists of a two-dimensional periodic square lattice of size $L^2$, occupied by $n$ volume-excluding particles. Each particle $\alpha\in \{1,\dots,n\}$ possesses a unit orientation vector $\mathbf{e}_\alpha$ that points toward one of the four neighboring sites. The orientation vector of each particle rotates $\pi/2$ clockwise or counter-clockwise with rate $D$. A particle moves to a vacant adjacent lattice site with rate $v_+$ if it points toward that lattice site, and with rate $v_0$ otherwise. The steady state of this model depends on the particle density $\phi = n/L^2$. At small values of $\phi$, typical configurations consist of small clusters of particles. Upon increasing $\phi$, for sufficiently large $v_+$, the system undergoes the nonequilibrium phase transition called motility-induced phase separation (MIPS). We shall show that the existence of this phase transition can be deduced by observation of a single trajectory obtained at a value of $\phi$ at which MIPS is not present.

{\em Training synthetic dynamics ---}  We introduce a general synthetic dynamics using a neural-network architecture called a transformer\c{vaswani2017attention}. We allow the transformer to know only that the dynamics is time-independent and consists of single-particle translations and rotations, although these restrictions can be lifted within this framework~\footnote{With other neural-network architectures these assumptions may also be lifted: time could be used as an additional input to the neural network, and collective updates could be achieved using an encoder-decoder architecture as used in language translation\c{vaswani2017attention}.}. In microstate $\ce$, the transformer represents the transition rates $W^{(\th)}_{\ce \to \ce'}$ to each of the microstates $\ce'$ connected to $\ce$ through translation or rotation of a single particle (\f{fig:schem}(c)).
The transformer learns which particle interactions are relevant to each of these moves, and what their rates are. To train the transformer we perform gradient ascent on its weights using backpropagation in order to maximize the log-likelihood $U_\omega^{(\th)}$, \eqq{phi}, with which it would have generated $\omega$. This trajectory is of length $T = 5\times 10^3$, on a $30 \times 30$ lattice, with parameters $\phi =  0.12\overline{4}, v_+ = 10, v_0 = 1$, and $D = 0.1$. MIPS is not present at these parameter values; see \f{fig:trajs_phi}.

During training we operate the transformer in one of two modes. In Mode 1, the transformer freely predicts $\ln W^{(\th)}_{\ce \to \ce'}$ for each possible transition. In Mode 2, the transformer assigns each transition to one of an integer number $N_W^{(\th)}$ of classes, and a second neural network assigns a value $\ln W^{(\th)}_{\ce \to \ce'}$ to each class. $N_W^{(\th)}$ is a hyperparameter that constrains the complexity of the learned dynamics, and provides a measure of the number of distinct classes of move (or processes) present in the original dynamics: the maximum value of  $U_\omega^{(\th)}$ obtained under training increases with $N_W^{(\th)}$ up to a value $N_W^{\star}$. The value $N_W^{\star}$ provides insight into the structure of the generator of the original dynamics, signaling, for instance, the presence of translational invariance. \s{transformer} provides additional details of the architectures of both types of neural-network dynamics and their optimization. We have used lattice models in this paper, but the transformer architecture can be directly applied to off-lattice models in any dimension.

In \f{fig:ll}(a) we show the results of training in Mode 1. The trajectory log-likelihood $U_\omega^{(\th)}$ increases with the number of observations (epochs) of the trajectory $\omega$, and converges to the value $U_\omega^{\star}$ that is obtained using the original dynamics. This value, not available to the transformer during training, indicates that the learned transition rates $W^{(\th)}$ are numerically very close to those of the original dynamics, $W^\star$. In \f{fig:ll}(b) we show the results of training in Mode 2, for several values of $N_W^{(\th)}$. These results show that $N_W^{\star}=4$, indicating that the transformer has correctly learned the degree of complexity of the original model, whose dynamical rules are translationally invariant and consist of 4 distinct rates. The inset to \f{fig:ll}(b) shows the evolution with training time of the values of the 4 rates, compared with their values in the original model.

During training we did not assume that the dynamical rules are local, nor that some processes (those that violate volume exclusion) are suppressed. The transformer was able to learn both things. If we know that interactions are of finite range then such knowledge can be used to reduce the number of transformer parameters required to learn dynamics (\s{fa_nn}). Transformers can also learn long-ranged interactions if they are present (\s{fa_cond}). We also note that learned rates for forbidden processes (inset  \f{fig:ll}(b)) are small and decrease with training time, but are not exactly zero: the result is that in forward-propagated trajectories a small fraction of particles can experience overlaps. If volume exclusion is suspected then it can be imposed directly. In addition, with Monte Carlo methods it is possible to determine that the rate of a forbidden process is exactly zero, even given a finite-length training trajectory; see \s{fa}.

{\em Discovering a phase transition using learned dynamics ---} In \f{fig:trajs_phi} we show that trajectories generated by the trained transformer can be used to determine the existence of a nonequilibrium phase transition not seen during training. We randomly initialize a configuration at a chosen density $\phi$ and propagate the transformer dynamics for fixed time $T$ (\f{fig:schem}(d)). At the training density $\phi= 0.12\overline{4}$, the model's steady state consists of small clusters, but trajectories generated by the transformer at larger values of $\phi$ show motility-induced phase separation: the transformer has therefore ``discovered'' this emergent phenomenon. 

In \f{fig:obs} we quantify the details of this phase separation. We measure the fraction of particles with four neighboring occupied sites $f_4$, and the variance of that quantity, as well as the number of clusters $n_{\rm c}$ and the average cluster size $s_{\rm c}$. The time averages of these observables are shown as a function of $\phi$ for trajectories obtained with the transformer, both in Mode 1 and Mode 2. For comparison, we show the same quantities from trajectories generated using the original dynamics. The agreement between original and learned dynamics is good, and slightly better using Mode 2, indicating that the transformer, trained under conditions for which no phase separation is observed (see the vertical line in the figure), has predicted the existence and details of a non-equilibrium phase transition.

{\em Conclusions ---} We have shown that the stochastic dynamics of a many-body system can be efficiently determined using machine-learning tools developed for language processing. A neural network called a transformer can function as an expressive ansatz for the generator of a many-body dynamics, for systems large enough that its possible rates are too numerous to represent explicitly. For instance, for the lattice model of active matter considered here, a $30 \times 30$ lattice at density $\phi=0.1$ admits $\binom{900}{90} \sim 10^{125}$ arrangements of particles. Each particle takes 1 of 4 rotational states, can move in 4 directions and undergo 2 types of rotation, meaning that there are in principle $\sim 10^{182}$ possible rates. Trained on this model, the transformer learns its dynamics, correctly identifying its local and translationally-invariant nature, and the numerical values of the associated rates. Forward-propagated trajectories of the transformer, carried out at higher densities than that observed during training, show motility-induced phase separation. The details of this nonequilibrium phase transition ``discovered'' by the transformer agree with those of the original model. Our work shows that it is possible to learn the dynamical rules of stochastic systems without explicit enumeration of rates or coarse-graining of configuration space, complementing existing papers on learning dynamics and pointing the way to the treatment of large and complex systems. 

{\em Acknowledgments} -- This work was performed as part of a user project at the Molecular Foundry, Lawrence Berkeley National Laboratory, supported by the Office of Science, Office of Basic Energy Sciences, of the U.S. Department of Energy under Contract No. DE-AC02--05CH11231. I.T. acknowledges NSERC. The computational resources (Stevin Supercomputer Infrastructure) and services used in this work were provided by the VSC (Flemish Supercomputer Center), funded by Ghent University, FWO and the Flemish Government – department EWI, and the National Energy Research Scientific Computing Center (NERSC), a U.S. Department of Energy Office of Science User Facility operated under Contract No. DE-AC02-05CH11231.\\

A tutorial for training the transformer can be found in Ref.~\footnote{\url{https://github.com/reproducible-science/learningDynamics}}.

\bibliography{bib}
\clearpage
\renewcommand{\theequation}{S\arabic{equation}}
\renewcommand{\thefigure}{S\arabic{figure}}
\renewcommand{\thesection}{S\arabic{section}}

\setcounter{equation}{0}
\setcounter{section}{0}
\setcounter{figure}{0}

\section{Path weight of a continuous-time Monte Carlo dynamics}
\label{ctmc}
\begin{figure*}[] 
   \centering
\includegraphics[width=\linewidth]{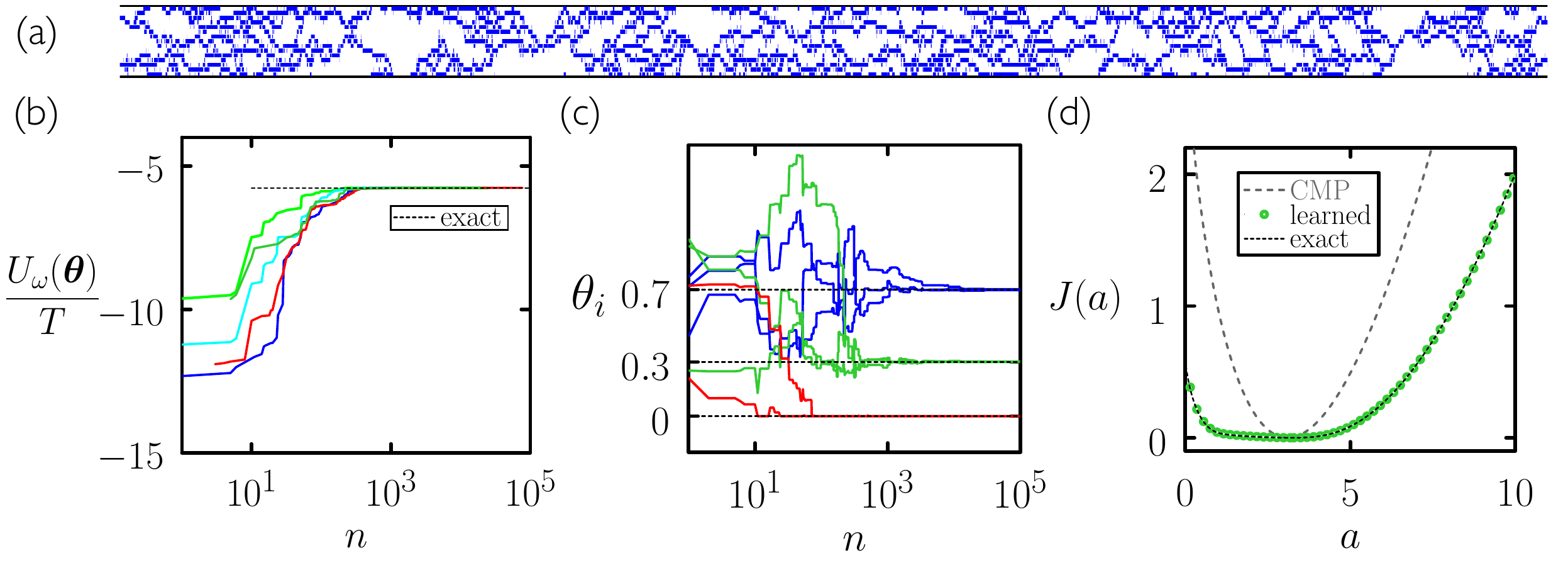} 
   \caption{\label{fig1} (a) Portion of the FA model trajectory $\omega$ from which we attempt to learn the FA model dynamical rules. Space is vertical, time is horizontal, and \texttt{1}s and \texttt{0}s are blue and white, respectively. (b) Log-likelihood (\eqq{phi2} divided by trajectory length $T$) with which 5 distinct trained synthetic models realize the trajectory shown in panel (a), as a function of the number of training steps $n$. (c) Rates for one of the trained synthetic dynamics as a function of $n$. Rates are color-coded according to their values in the original model: blue, green, and red rates have values in the original model of 0.7, 0.3, and 0, respectively. (d) The large-deviation rate function calculated from forward propagation of 100 distinct trained synthetic models (green) compared with the rate function of the true FA model (black).}
\end{figure*}

Consider a dynamical trajectory $\omega$ of total time $T$. The trajectory starts in configuration (microstate) $\ce_0$, and visits $K$ additional configurations $\ce_k$. In configuration $\ce_k$ it is resident for time $\Delta t_{C_k}$. Schematically, 
\beq
\omega = \ce_0 \xrightarrow[]{\Delta t_{\ce_0}} \ce_1 \xrightarrow[]{\Delta t_{\ce_1}} \cdots\ce_{K-1} \xrightarrow[]{\Delta t_{\ce_{K-1}}}\ce_K \xrightarrow[]{\Delta t_{K}} \ce_K, \nonumber
\eeq
where $\Delta t_K \equiv T-\sum_{k=0}^{K-1} \Delta t_{\ce_k}$.

We are told that $\omega$ was generated by a continuous-time Monte Carlo dynamics\c{gillespie1977exact,gillespie2007stochastic} whose rates $W^{\star}_{\ce \to \ce' }$ for passing between configurations $\ce$ and $\ce'$ we do not know. We will call this unknown dynamics the {\em original} dynamics.

In order to learn the original dynamics we introduce a new continuous-time Monte Carlo model called the {\em synthetic} dynamics. The synthetic dynamics consists of a set of allowed configuration changes $\{\ce \to \ce'\}$, which must include those observed in $\omega$, and associated rates $W^{(\th)}_{\ce \to \ce' }$. Rates are parameterized by a vector $\th = \{\theta_1,\dots,\theta_N\}$ of $N$ numbers (in the main text these numbers corresponds to the weights of the transformer).

We train the synthetic dynamics by maximizing the log-likelihood $U_\omega^{(\th)}$ with which it would have generated $\omega$. To calculate $U_\omega^{(\th)}$ we start by considering the portion 
\beq
\ce_k \xrightarrow[]{\Delta t_{\ce_k}} \ce_{k+1}
\eeq
of $\omega$, which involves a transition $\ce_k \to \ce_{k+1}$ and a residence time $\Delta t_{\ce_k}$. The probability with which the synthetic dynamics would have generated the transition $\ce_k \to \ce_{k+1}$ is 
\beq
W_{\ce_k \to \ce_{k+1}}^{(\th)}/R_{\ce_k}^{(\th)},
\eeq
where $R_{\ce_k}^{(\th)} \equiv \sum_{\ce'} W_{\ce_k \to \ce'}^{(\th)}$, the sum running over all transitions allowed from $\ce_k$. The probability density~\footnote{Working with the probability $R_{\ce_k}^{(\th)} \e^{-\Delta t_{\ce_k} R_{\ce_k}^{(\th)}} \Delta t_{\ce_k}$ gives rise to an additional term $\sum_{k=0}^{K-1} \Delta t_{\ce_k}$ in \eq{phi2} that does not depend on the choice of synthetic dynamics and may be omitted without consequence.} with which the synthetic dynamics would have chosen the associated residence time $\Delta t_{\ce_k}$ is 
\beq
R_{\ce_k}^{(\th)} \e^{-\Delta t_{\ce_k} R_{\ce_k}^{(\th)}}.
\eeq
The product of transition- and residence-time factors is 
\beq
W_ {\ce_k \to \ce_{k+1}} ^{(\th)}\e^{-\Delta t_\ce R_{\ce_k}^{(\th)}} \equiv p_{\ce_k}.
\eeq
Noting that the probability of the final portion of the trajectory, $\ce_K \xrightarrow[]{\Delta t_{K}} \ce_K$, is 
 \beq
1- \int_0^{\Delta t_K} {\rm d} \tau \, R^{(\th)}_{\ce_k} \e^{-R^{(\th)}_{\ce_k}\tau} = \e^{-\Delta t_K R^{(\th)}_{\ce_K}} \equiv p_K,
 \eeq
the log-likelihood with which the synthetic dynamics would have generated $\omega$ is 
 \bea
\label{phi2}
U_\omega^{(\th)}&=&\ln  \left( p_K \prod_{k=0}^{K-1} p_{\ce_k} \right) \nonumber \\
&=&\sum_{k=0}^{K-1} \left(\ln W_{\ce_k \to \ce_{k+1}}^{(\th)} -\Delta t_{\ce_k} R_{\ce_k}^{(\th)}\right) \nonumber \\
&-& \Delta t_K R_{\ce_K}^{(\th)}.
\eea
The sum in \eq{phi2} is taken over the trajectory $\omega$, i.e. over all configuration changes and corresponding residence times. To train the synthetic dynamics we adjust its parameters $\th$ until \eq{phi2} no longer increases. The synthetic dynamics obtained in this way -- the {\em learned} dynamics -- is then the best approximation to the original dynamics that our choice of allowed configuration changes and method of training allows: for a sufficiently long trajectory $\omega$, the dynamics that maximizes \eq{phi2} is the original dynamics $W^{\star}_{\ce \to \ce' }$.

In sections \ref{fa}, \ref{fa_nn} and \ref{fa_cond}, we illustrate the training procedure in situations of increasing complexity.

\section{Transformer and training details}
\label{transformer}

The neural network used to treat the active-matter model described in the main text (and the FA models described in \s{fa_nn} and \s{fa_cond}) is a transformer\c{vaswani2017attention}. A transformer relies on an attention mechanism that allows it to learn which parts of a configuration are relevant for a particular process. This generality ensures that it is not biased toward learning local interactions, as is the case for e.g. a convolutional neural network.

The first step in calculating the transition rates is a learned representation of the current state of the system.
We first embed particle positions and orientations (or spin states for the FA model) as $d_h$-dimensional vectors using trainable weight matrices; $d_h$ is a hyperparameter controlling the expressivity of our neural-net model. We then sum the representations of the position and spin for each particle, which serve as the input to the transformer. We do not impose the boundary conditions of our lattice models; the transformer has to learn these through its positional embedding. For the lattice active matter model, we do not use the empty sites for computational efficiency.  Instead, the transformer must learn which neighboring sites are occupied.

Next, we calculate the attention matrix for the entire configuration using multi-head attention\c{vaswani2017attention}, and for each particle obtain a $d_h$-dimensional vector containing a weighted sum of the properties of all other particle (the weighting being a measure of the attention paid to each particle).
These vectors are then processed using fully-connected neural networks.
We apply this alternating process of attention and application of fully-connected neural networks $n_l$ times.
The final output of these blocks is used to calculate the transition rate for each possible particle update (spin flip for the FA model or particle rotation or translation for the active-matter model).

Training in Mode 1, the rates are obtained by applying a fully-connected neural network to the output vectors of the transformer. We apply the same network for each particle. This fully-connected neural network has one output node for each possible particle update, the value of $\ln W$ assigned to the corresponding transition. Training in Mode 2, we first classify the transformer's output vectors using a fully-connected neural network with $N_W$ output nodes and a softmax activation function, again for each possible particle update. The class with the highest probability is sent, as a one-hot vector, to another fully-connected neural network with one output node, which calculates the value of $\ln W$ for each of the $N_W$ classes. Picking the highest-probability class is not a differentiable operation, and so we use a straight-through estimator to obtain the gradients to optimize these neural networks\c{jang2016categorical} .

The results in this paper were obtained with the hyperparameters $d_h = 64$ and $n_l =2$. We used the AdaBelief optimizer\c{zhuang2020adabelief} with a learning rate of $10^{-4}$ to optimize the transformer's weights.
To obtain a baseline for the trajectory log-likelihood $U_\omega^{(\th)}$, we first train a Mode 1 neural-network dynamics on the provided trajectory. For efficiency we train for several epochs on smaller sections of the trajectory; during the final stages of training we use the entire trajectory to obtain more accurate gradients of the trajectory log-likelihood.
Next, we train a Mode 2 neural-network dynamics to gain insight into the model's generator. We initialize the first layers of the neural network (the embedding and transformer layers) with the weights obtained with the Mode 1 dynamics, which leads to much faster convergence.\\

\section{Learning a local dynamics with knowledge of its locality}
\label{fa}

In this section we consider an original dynamics whose rules are local, and assume that we know that its rules are local. The learning procedure then amounts to identifying the correct numerical values for the rates of each local process. This is a conceptually simple case, but worth considering because it illustrates the precision with which rates can be learned, the fact that it is possible to learn the existence of forbidden processes, and also demonstrates some of the convenient features of learning dynamics offline, without propagating new trajectories.  

We consider the original dynamics to be the one-dimensional Fredrickson-Andersen (FA) model with periodic boundary conditions\c{fredrickson1984kinetic}. The FA model is a lattice model of a supercooled liquid whose dynamical rules give rise to slow relaxation and complex space-time behavior\c{butler1991origin,garrahan2002geometrical}. On each site of a lattice lives a binary spin that can be down (\texttt{0}) or up (\texttt{1}), intended to model immobile and mobile regions of a supercooled liquid. The dynamical rules for the model are nearest-neighbor ones: spins can only flip if at least one of their nearest neighbors is up; if so, down spins flip up with rate $c$, and up spins flip down with rate $1-c$. We choose $c=0.3$, giving the rates shown in \t{tab1}. We use the shorthand \1, \5, etc. to denote the 8 possible configuration changes of a nearest-neighbor dynamics: each triplet indicates the process in which the central spin changes state. Thus \3 indicates the process \3 $\to$ \1, while \0 indicates the process \0 $\to$ \2, etc. The rates for the processes \0 and \2 in the FA model are zero, a feature responsible for the model's complex dynamical behavior. 

\begin{table}[h!]
  \begin{center}
    \begin{tabular}{c|c|c} 
     process & true rate & learned rate \\
      \hline
\0  &0 &0\\
 \1  &0.3 & 0.302028\\
 \2 &0 &0 \\
 \3  &0.7 & 0.699833\\
 \4 &0.3 & 0.300031 \\
 \5 &0.3 & 0.301362 \\
 \6 &0.7 & 0.697800 \\
 \7  &0.7 & 0.702803
    \end{tabular}
    \caption{\label{tab1} Comparison of the true FA model rates and those learned from the trajectory shown in \f{fig1}(a).}
 \end{center}
\end{table}

We start by generating a single FA model trajectory $\omega$ of length $T=10^6$, using a model with 15 lattice sites. This is the trajectory from which we attempt to learn the rates of the model that generated it. A segment of this trajectory of length $T/500$ is shown in \f{fig1}(a). 

To construct the synthetic dynamics we make the assumption that only local processes are allowed, i.e. that only one spin at a time can flip. We also assume that the dynamical rules of the model are independent of time. In this section we further assume that the dynamical rules are local, and that these rules are translationally invariant. The synthetic dynamics constrained in this way contains 8 parameters $\th=\{\theta_1,\dots,\theta_i,\dots,\theta_8\}$ that correspond to the rates of the 8 processes shown in~\t{tab1}.
\begin{figure}[] 
   \centering
\includegraphics[width=0.7\linewidth]{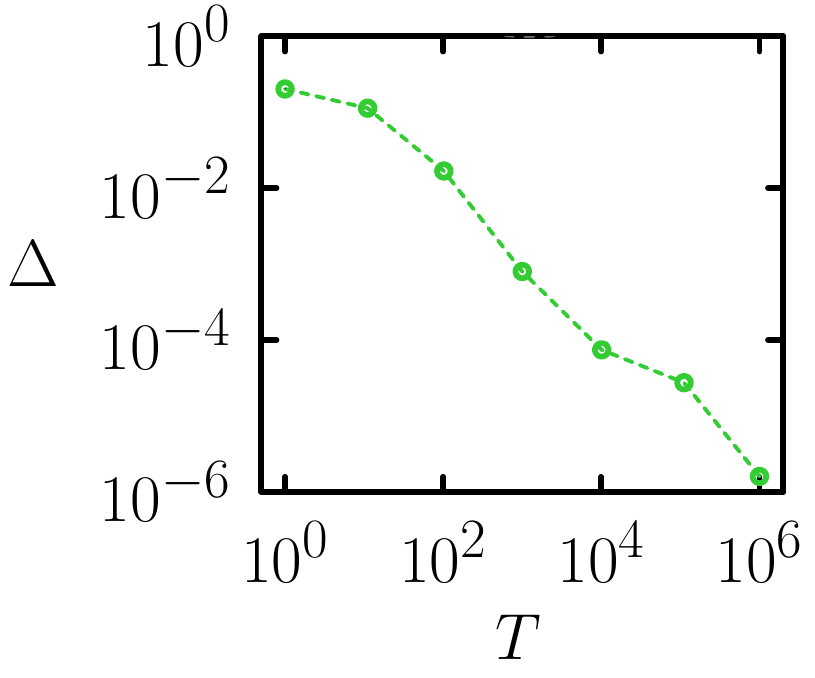} 
   \caption{\label{fig1b} A measure of the error between original and learned rates, \eqq{delta}, as a function of the length $T$ of the training trajectory $\omega$.}
\end{figure}

To find the rates that maximize \eq{phi2} we proceed as follows. We initialize the rates $\theta_i$ by choosing them to be random numbers uniformly distributed on $(0,1]$, and then apply the following Monte Carlo algorithm. At each step of the learning procedure we propose new parameters
\beq
\theta_i \to \max\left(0,\theta_i +{\cal N}(0,\sigma^2) \right),
\eeq
and accept the proposal if \eq{phi2} increases or remains the same. This Monte Carlo algorithm is equivalent, for small values of the proposal-size parameter $\sigma$, to noisy clipped gradient ascent on the function $U_\omega^{(\th)}$, described (for $\theta_i>0$) by the Langevin equation\c{whitelam2020correspondence}
\bea
\label{lang2}
\frac{\d \theta_i}{\d n} = \frac{\sigma}{\sqrt{2 \pi}}\frac{1}{|\nabla U_\omega^{(\th)}|}\frac{\partial U_\omega^{(\th)}}{\partial \theta_i}+ \eta_i(n).
\eea
Here $n$ is the number of steps of the learning algorithm, $\nabla$ is the $N$-dimensional gradient in the coordinates $\th$, and $\eta$ is a Gaussian white noise with zero mean and covariance $\av{\eta_i(n)\eta_j(n')}=\sigma^2\delta_{ij} \delta(n-n')/2$. We set $\sigma=0.05$.

We carried out 100 independent learning simulations, each begun from different random initial rates $\theta_i$, and each trained on the single trajectory $\omega$. Each simulation converged, within about $10^3$ steps, to the same value of \eq{phi2}; we show 5 examples in \f{fig1}(b) (colored lines). This value is equal to \eq{phi2} evaluated using the true rates of the FA model (black dashed line), although that information was not available to the algorithm during training. The rates produced by one learning simulation are shown in \f{fig1}(c) and \t{tab1}: the learned rates are numerically close to those of the FA model.
\begin{figure}[] 
   \centering
\includegraphics[width=0.9\linewidth]{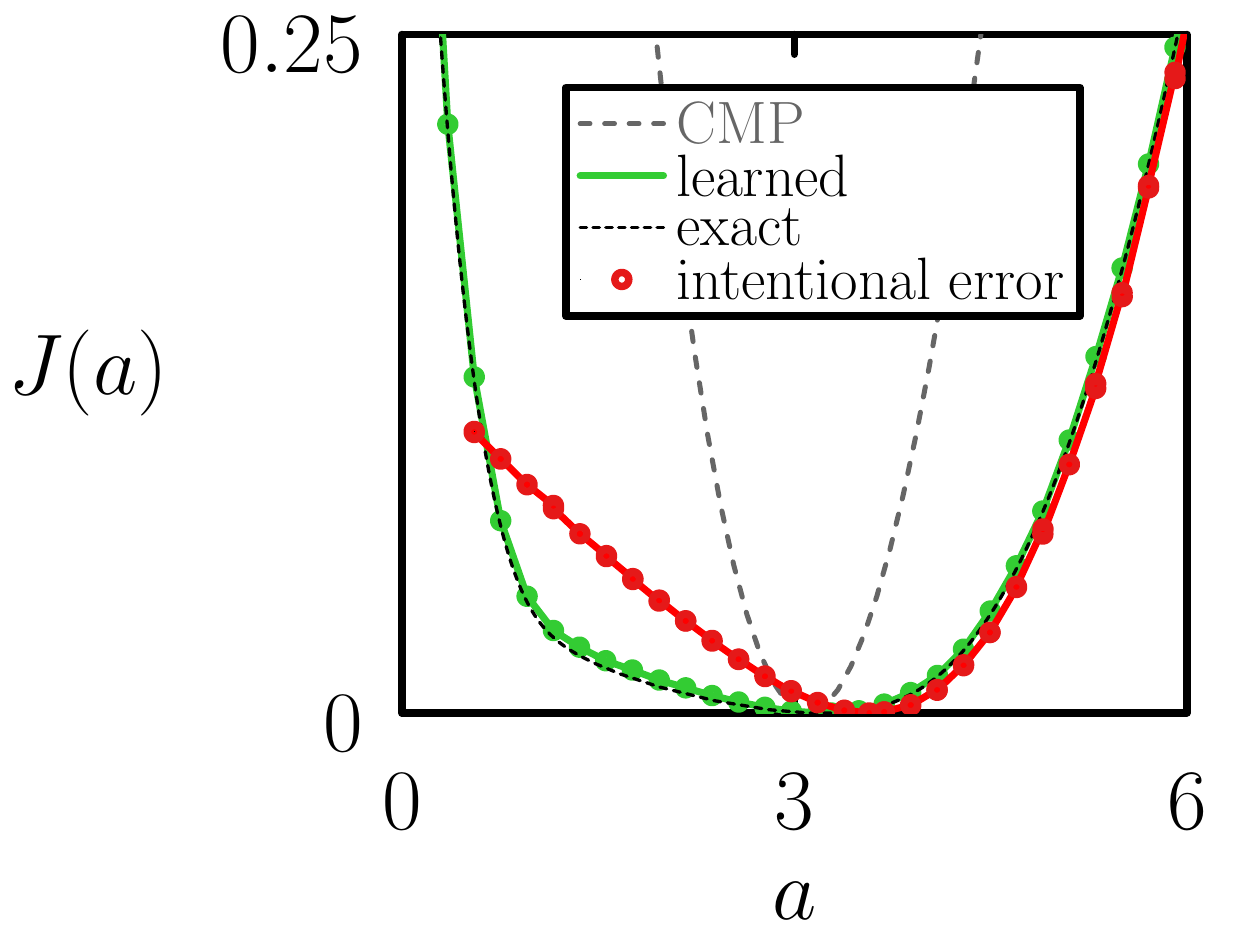} 
   \caption{\label{fig3} As \f{fig1}(d), but now the red curve results from calculations in which the value $10^{-2}$ was added to the learned rates \0 and \2 in order to introduce an intentional error. The result -- substantially different to the exact answer -- shows that comparing the fluctuations of learned and true dynamics is a discriminating test of the learning process, and highlights the precision of that process.}
\end{figure}
\begin{figure*}[] 
   \centering
\includegraphics[width=0.9\linewidth]{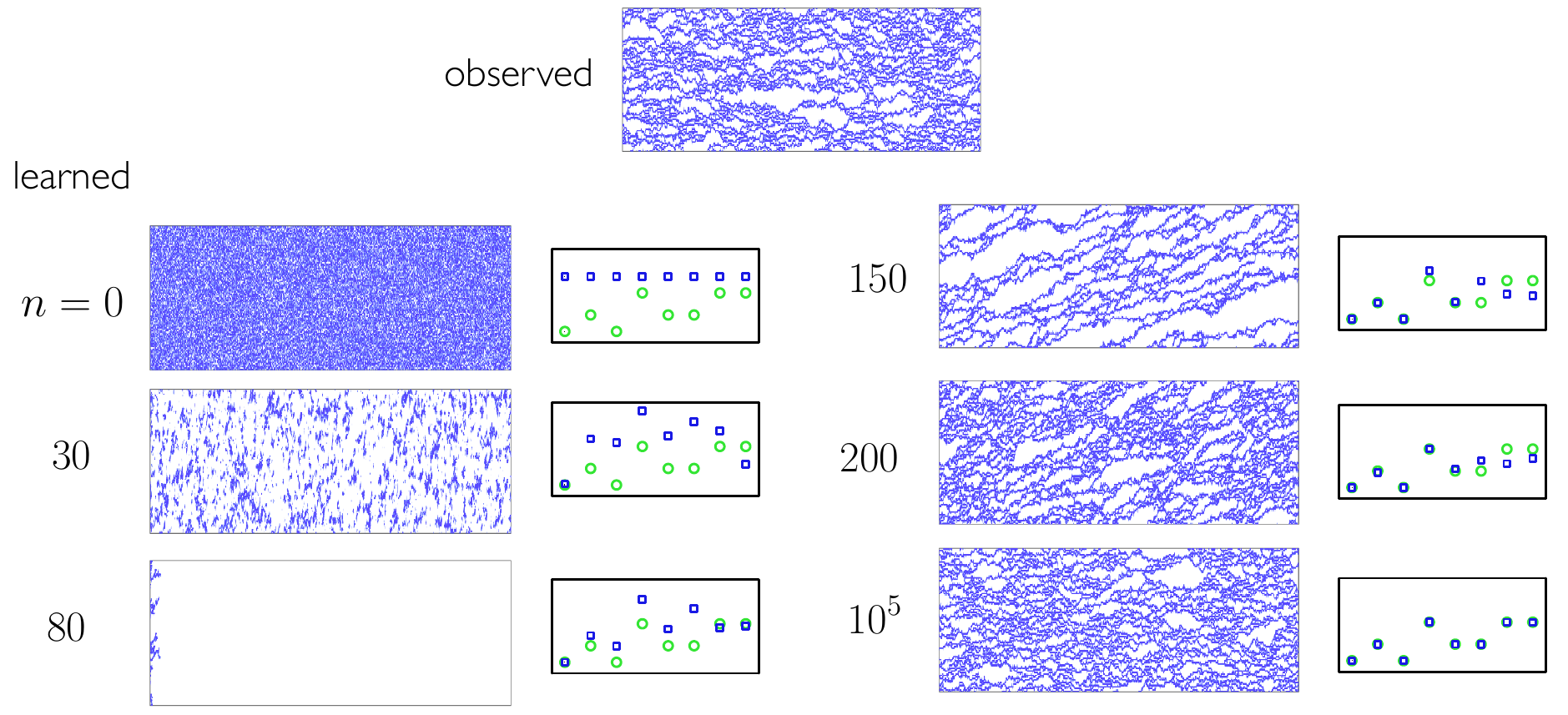} 
   \caption{\label{fig2} Comparison of the true FA model dynamics (top) with the forward-propagated dynamics of the synthetic models produced after $n$ steps of training. Small panels show the 8 learned rates (blue squares) and true rates (green circles).}
\end{figure*}

Notably, the learning process has correctly identified that the rates \0 and \2 of the model used to produce the trajectory of \f{fig1}(a) are exactly zero. Observing that neither transition occurs in a trajectory of finite length allows us only to bound the rates with which those processes occur. The learning procedure has done better, determining that vanishing rates lead to the largest attainable value of \eq{phi2}, and has therefore identified the existence of forbidden processes in the FA model.

The non-vanishing rates shown in \f{fig1}(c) and \t{tab1} are close to those of the FA model, but not identical. A natural question to ask is how well the learned models can approximate the dynamics generated by the original one. Doing so requires the introduction of order parameters. A popular order parameter for the FA model is the {\em activity} $a=A/T$, the number of configuration changes $A$ per unit time $T$\c{bodineau2012finite,garrahan2016classical}. Forward-propagated trajectories of duration $T=10^6$ of the learned models indeed have the same typical activity as the FA model trajectory, $a_0 \approx 3.2$. 

However, a more discriminating measure of similarity is to compare the fluctuations of the learned and original dynamics. Fluctuations of the activity can be characterized by the dynamical large-deviation rate function $J(a) = \lim_{T \to \infty} -T^{-1} \ln \rho_T(A)$, a measure of the logarithmic probability $\ln \rho_T(A)$ of observing, for a trajectory of length $T$, a particular value of $A$\c{den2008large,touchette2009large}. We used the VARD method described in Refs.\c{whitelam2019evolutionary,jacobson2019direct} to calculate $J(a)$ for the learned dynamics (we used the neural-network ansatz described in\c{whitelam2019evolutionary}), with each of the 100 learned models constrained to produce a different value of $a$. Values of $J(a)$ calculated in this way are shown as green circles in \f{fig1}(d); they closely approximate the exact rate function of the FA model (black dashed line), which we calculated by numerically diagonalizing the model's rate matrix\c{touchette2009large}.

The numerical similarity between $J(a)$ of the original and learned models indicates that the precision indicated in \t{tab1} and in panels (a) and (b) of \f{fig1} is sufficient to produce an essentially exact description of the behavior of the original model, at least as far as the activity is concerned. (In \f{fig3} we show results in which we have made an intentional error by adding $10^{-2}$ to the learned rates \0 and \2; in that case there is a clear discrepancy between the learned and true rate functions.) 

This comparison also indicates that there is sufficient information in $\omega$ to calculate the likelihood with which the original dynamics would produce trajectories never seen in $\omega$. The Conway-Maxwell-Poisson (CMP) function shown in gray in \f{fig1}(d) is an upper bound on $J(a)$ inferred by sampling trajectories containing {\em only} typical configurations\c{garrahan2017simple,whitelam2019evolutionary}, such as those seen in \f{fig1}(a). The large discrepancy between the CMP bound and the true rate function indicates that rare trajectories of the FA model are dominated by configurations very different to those seen in $\omega$, and therefore never observed during training. Nonetheless, rates inferred from observation of $\omega$ can be used to calculate the probability with which long trajectories containing these previously unseen, rare configurations will be observed.

The precision of learning increases with the length of the trajectory $\omega$. In \f{fig1b} we show 
\beq
\label{delta}
\Delta \equiv N^{-1} \sum_{i=1}^N \left(\theta_i -\theta_i^{\star} \right)^2,
\eeq
the mean-squared difference between the $N=8$ rates of the FA model $\theta_i^{\star}$ and the rates $\theta_i$ learned from an FA model trajectory $\omega$ of length $T$ (a single trajectory is used for each value of $T$). The precision of the learning process increases with $T$ over the range shown.

We end this section by highlighting a notable feature of the ``offline'' learning process. During training, no trajectories of the synthetic dynamics are propagated. Given a synthetic dynamics we propose a change of its parameters, calculate the likelihood with which that dynamics {\em would} have generated the original trajectory $\omega$, and accept the proposal if \eq{phi2} does not decrease. Some of the proposals accepted in this way would have been rejected had we trained by comparing $\omega$ with forward-propagated trajectories of the synthetic dynamics. In \f{fig2} we propagate some of the synthetic models produced after $n$ steps of training (here we consider an FA model of 100 sites). Next to each trajectory we display a panel comparing the learned rates (blue squares) and true rates (green circles). Here, synthetic rates were initially set to unity. As with the smaller FA model, the training procedure converges to a good approximation of the true rates, and again the rates of the forbidden transitions \0 and \2 are correctly identified to be zero. In general terms the synthetic trajectories look increasingly like those of the FA model as $n$ increases, but there are some notable exceptions. At step $n=80$ we encounter an absorbing state of all \texttt{0}s. This configuration is not accessible to the FA model from any configuration with at least one up spin, and if we were to train by comparison of trajectories generated by original- and synthetic dynamics then this synthetic dynamics would be rejected. However, such a comparison is never made during offline training, and after additional training steps the synthetic dynamics converges to the true dynamics. 

\section{Learning a local dynamics {\em without} knowledge of its locality}
\label{fa_nn}

In \s{fa} we assumed that $\omega$ was generated by a dynamics whose rules are local and translationally invariant, allowing us to define a synthetic dynamics with few parameters. If we relax these restrictions then the number of possible rates increases exponentially with system size, and so direct representation of the rates of the synthetic dynamics becomes impractical for large systems. Instead, we can express a general synthetic dynamics using a neural network, which here we take to be a transformer. 

We now relax the assumptions of locality and translational invariance in its interaction rules, and so the transformer must determine which features of the configuration are relevant to each process (here we use the version of the FA model whose rates are proportional to the number of nearest neighbors in the up state). We generate a training trajectory $\omega$ of length $T = 10^6$ using a model with $N=100$ lattice sites and rate parameter $c=0.3$. For each configuration of the trajectory the transformer must calculate the rate of flipping each spin, and therefore has to represent $N \times 2^N$ possible transition rates.  

We first trained a transformer in Mode 1 (making no restrictions on the number of distinct rate values), and observed that the transformer has the capacity to represent the transition rates for this large state space: the trajectory log-likelihood obtained with the synthetic dynamics rapidly converges to that obtained with the original dynamics. To gain further insight into the generator of the observed trajectory we trained a transformer in Mode 2 on the same trajectory. In Mode 2, the number of distinct rates is limited to $N_W^{(\th)}$, a model hyperparameter. The transformer now must assign each transition to one of $N_W^{(\th)}$ classes, and a second neural network determines the rate for each of these classes.
\begin{figure}[] 
\includegraphics[width=\columnwidth]{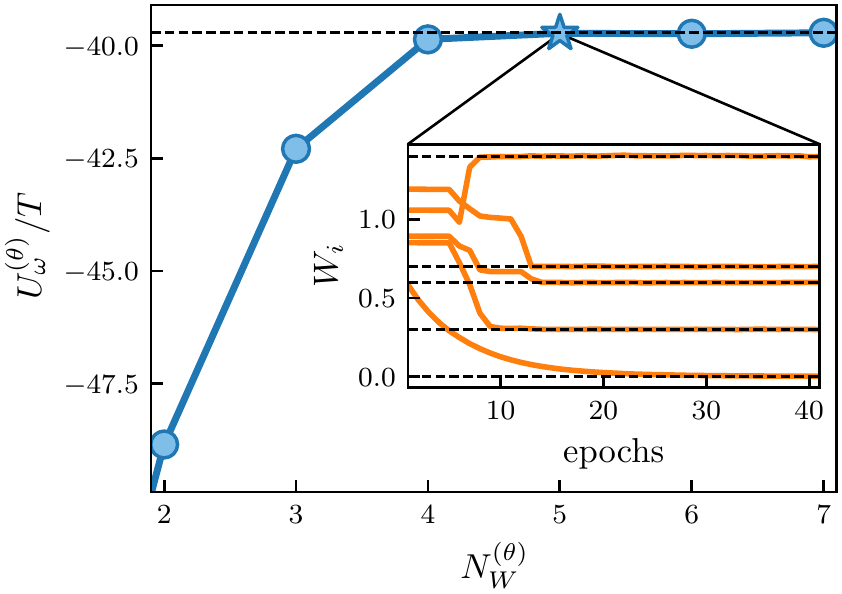} 
   \caption{\label{fig:fa}  Training a transformer in Mode 2 to maximize the log-likelihood $U_\omega^{(\th)}$ with which it would have generated a trajectory of an FA model with 100 lattice sites. The value of $U_\omega^{(\th)}$ is maximized if $N_W^{\th} \ge 5$, which is equal to the number of distinct rates of the original model. The horizontal black line denotes the exact value. Inset: Evolution of the rates during training, for the case $N_W^{(\th)} = 5$. The horizontal black lines denote the values of the rates in the original dynamics.}
\end{figure}

In \f{fig:fa}, we show how the optimized trajectory log-likelihood depends on the value of  $N_W^{(\th)}$. As the number of distinct rates is increased, the trajectory log-likelihood increases until $N_W^{(\th)} \ge 5$, at which point it remains constant. The small number of rates required to model these dynamics tells us that the original dynamics is translationally invariant. We also gain insight into the number of distinct processes present in the original dynamics (because multiple processes can have the same transition rate, this value is a lower bound on the number of distinct processes). In this case, the value $N_W^\star = 5$ is equal to the number of rates in the original dynamics; we compare the exact and learned rates in the inset of \f{fig:fa}.

While restricting the interactions to each spin's nearest neighbors limited the number of rates in \s{fa} such that they can be modeled explicitly, prior knowledge of the interactions can also be built into our neural-network framework. For instance, if a maximum interaction range is assumed, local attention\c{beltagy2020longformer} can be used which limits the attention calculation to a fixed number of neighboring sites, reducing the computational cost and accelerating convergence.

\section{Learning a nonlocal dynamics}
\label{fa_cond}

The dynamical rules in the lattice active matter model described in the main text and the FA model described in the previous section are local, depending only on nearest-neighbor particles. In this section we demonstrate that the transformer can likewise learn nonlocal dynamical rules, where rates depend on the states of distant particles and the number of distinct rates is large. We consider a conditioned dynamics of the FA model, whose trajectories are biased towards having atypical values of the activity. The generator of this model is given by 
 \begin{equation}
    W_s^\text{doob}= \mathcal{L}[W_s - \theta(s) I]\mathcal{L}^{-1},
\end{equation}
where $W_s$ is obtained by multiplying the off-diagonal elements of the FA-model generator $W$ by $e^{-s}$, $\theta(s)$ is the largest eigenvalue of $W_s$,  and $\mathcal{L}$ is the corresponding left eigenvector of $W_s$ as a diagonal matrix\c{chetrite2014,causer2021optimal}.
Although the sets of forbidden and allowed transitions are the same for both $W$ and $W_s^\text{doob}$, the transition rates of $W_s^\text{doob}$ depend on the entire lattice configuration. Recently, the conditioned dynamics of several  prototypical dynamical systems have been uncovered using neural-network methods\c{casert2021dynamical, yan2022learning} and reinforcement learning\c{rose2021reinforcement, das2021reinforcement}.

In \f{fig:fa_cond}, we show the result of training a transformer in Mode 1 on a trajectories of conditioned FA models of length $T = 10^7$ with $N=14$ lattice sites.
As values for the conditioning field, we use $s = s^\star \approx 0.017$ for which the susceptibility $\chi(s)= \theta^{\prime\prime}(s)$ is maximal for this lattice size, and $s = -0.1$. Typical trajectories of the conditioned dynamics for $s= -0.1$ have larger activity than typical trajectories of the unconditioned FA model.
Trajectories generated by the trained transformers are shown in \ff{fig:fa_cond}(a) and (e); compare the unconditioned dynamics shown in \ff{fig1}(a).
The rapid convergence of $U_\omega^{(\th)}$ to its exact value for both values of $s$ is shown in \ff{fa_cond}(b) and (f), and demonstrates the transformer's ability to correctly represent the long-range interactions of the conditioned FA model. In \ff{fa_cond}(c) and (g), we further validate this statement by measuring the probability of the lattice containing $n$ spins in state \texttt{1} over trajectories created with the original dynamics and by by forward-propagating the transformer dynamics. In \ff{fa_cond}(d) and (h), we study the distribution of the activity for conditioned trajectories with both dynamics. The agreement with the exact conditioned dynamics is good.

\onecolumngrid

\begin{figure}[h] 
\includegraphics[width=\linewidth]{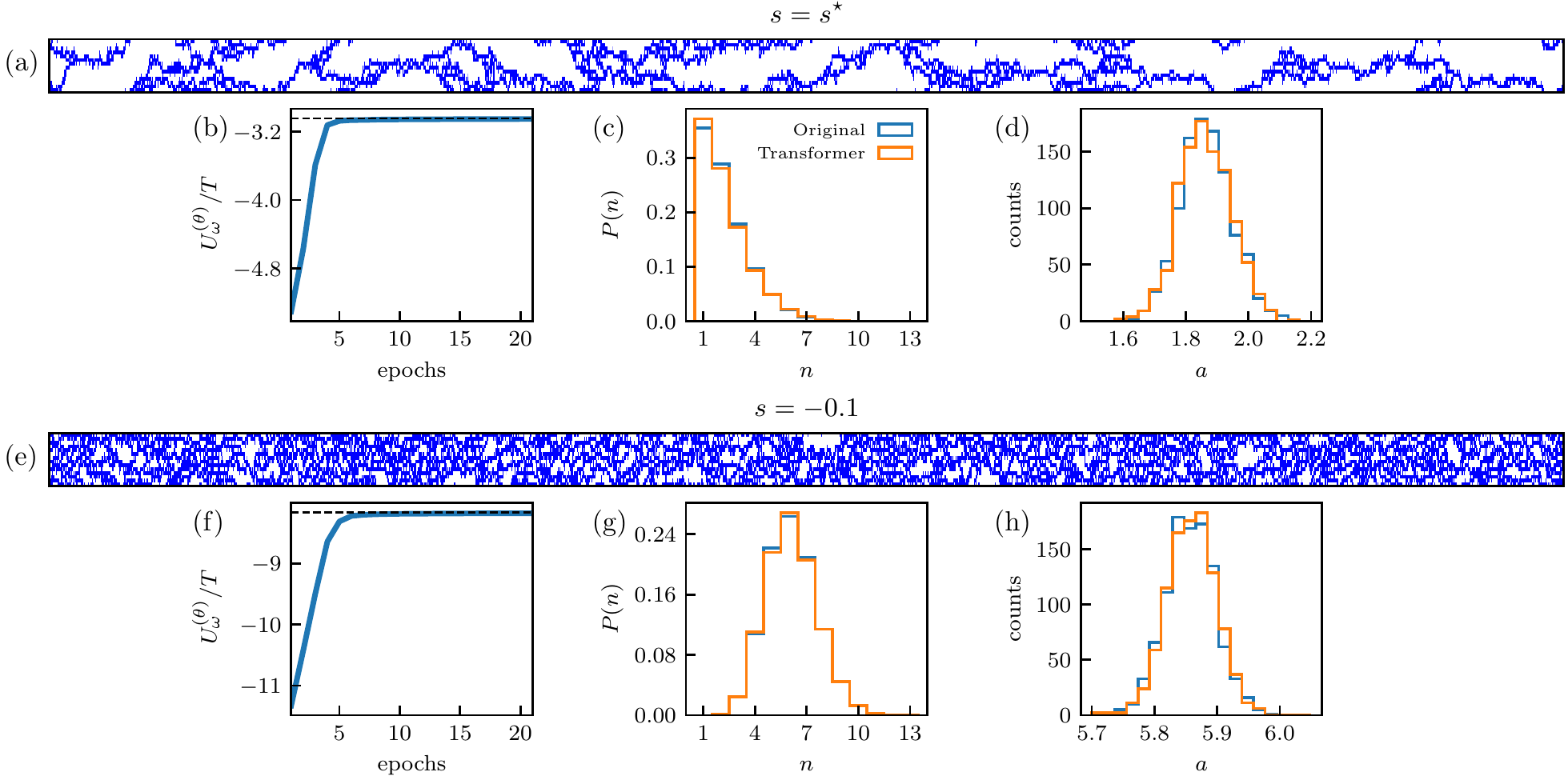} 
   \caption{\label{fig:fa_cond}Learning the long-ranged dynamics of the conditioned FA model. The dynamical rules of the conditioned generator are nonlocal and consist of a large number of distinct rates. (a) Trajectory of length $T= 2000$ generated by a transformer trained on a trajectory of the conditioned FA model with $N = 14$ lattice sites, $c= 0.3$, and $s = s^\star \approx 0.017$.  Space is vertical, time is horizontal, and \texttt{1}s and \texttt{0}s are blue and white, respectively. (b) Optimization of a transformer to maximize the log-likelihood of the training trajectory. The horizontal black line denotes the value of the path weight associated with the original model. (c) Probability of the lattice containing $n$ spins in state \texttt{1} during a trajectory of length $T= 10^6$, both for the original (exact conditioned) dynamics and the transformer dynamics. (d) Histogram of the activity $a$ of 1000 trajectories of length $T = 10^4$. (e--g) As (a--d), but now for $s= -0.1$. }
   \end{figure}

\end{document}